\documentclass[preprint,showpacs]{revtex4}
\usepackage{graphicx}
\usepackage{epsfig}
\usepackage{axodraw}
\begin{document}
\title{Neutron Charge Radius: Relativistic Effects and the Foldy Term}
\author{ W.R.B. de Ara\'ujo$^a$,
T. Frederico$^a$, M. Beyer$^b$, and H.J. Weber$^c$}
\address{$^a$ Dep. de F\'\i sica, Instituto Tecnol\'ogico de Aeron\'autica,
Centro T\'ecnico Aeroespacial, \\
12.228-900 S\~ao Jos\'e dos Campos, S\~ao Paulo, Brazil.}
\address{$^b$ Fachbereich Physik,
Universit\"at Rostock, 18051 Rostock, Germany}
\address{$^c$ Dept. of Physics, University of Virginia,
Charlottesville, VA 22904, U.S.A.}
\date{\today}
\begin{abstract}
The neutron charge radius is studied within a light-front model
with different spin coupling schemes and wave functions. The cancellation
of the contributions from the Foldy term and  Dirac form factor  to the neutron
charge form factor is verified for large nucleon sizes
and  it is independent of the detailed form of quark spin coupling
and wave function. For the physical nucleon our results for the
contribution of the Dirac form factor to the neutron radius are insensitive
to the form of the wave function while they strongly depend on the
the quark spin coupling scheme.
\end{abstract}
 \maketitle
\section{Introduction}

Nowadays, there is a renewed interest in the nucleon
electromagnetic form factors due to recent precise
experiments\cite{exp1}. The electroweak
static observables are receiving  theoretical attention as well,
in particular, the neutron charge radius was recently studied
with respect to its relativistic origin\cite{isgur,asfbw,simula0,karmanov02}.
{}From the nonrelativistic point of view the nucleon
wave function with SU(6) symmetry implies a zero neutron
charge radius and $\mu_p/\mu_n=-m_N/(2 m)$ ($m_N$ and $m$ are the masses of
the nucleon and constituent quarks, respectively). However, the
quark spin-spin interaction that implies the nucleon-delta
mass splitting allows the dynamical breaking of the SU(6) symmetry
and a nonvanishing neutron charge mean square radius ($r_n^2$).
The flavor identical
quarks are pushed out, while the $u$ quark stays around the neutron center,
which gives a negative charge mean square radius, in agreement with the
observed sign\cite{kop}.

The experimental
value $r_n^2=\ -0.113 \pm 0.005 \ fm^2$ \cite{kop} is coincidentally
near the contribution of the Foldy term
($ \frac32\frac{\mu_n}{m^2_n}=-0.126 \ fm^2$), and for
that reason the contribution of the Dirac form factor ($F_{1n}$)
is quite small, $r^2_{1n}= \ 0.013 \pm 0.005 \ fm^2$.  The electric
and magnetic form factors (Sachs form factors)  are given by:
\begin{eqnarray}
G_{EN}(q^2)&=& F_{1N}(q^2)+\frac{q^2}{4m_N^2}F_{2N}(q^2) \ ,
\nonumber \\
G_{MN}(q^2)&=& F_{1N}(q^2)+F_{2N}(q^2) \ ,
\end{eqnarray}
where $N= \ n$ or $p$, $F_{2N}$ is the Pauli form factor,
$q^\mu$ is the momentum transfer. The magnetic moment
is $\mu_N=G_{MN}(0)$ and the charge mean square
radius is $r^2_N=6\frac{dG_{EN}(q^2)}{dq^2}|_{q^2=0}$.

The naive physical picture of the spin-spin interaction dominating
the negative neutron charge radius is confronted with the fact
that the recoil effect from the Foldy term dominates the charge
radius. Thus, if one associates the intrinsic  charge distribution
with $F_{1n}$, one comes to the wrong conclusion that the
spin-spin interaction is not relevant for $r^2_n$. The recent work
of Isgur\cite{isgur} clarified this issue, the rest frame charge
distribution of the neutron should be associated with the form
factor $G_{En}$ and not with $F_{1n}$. He found that in the the
first relativistic correction to the Dirac form factor cancels the
Foldy term in a relativistic model of the nucleon within a
quark-diquark picture. Consequently, the nonzero value of the
neutron charge radius should reflect the intrinsic charge
distribution and the detailed quark spin dynamics. In fact, in
Ref.\cite{asfbw} it was shown that the different forms of
relativistic quark spin coupling in the nucleon have dramatic
effects on the neutron charge mean square radius. The relativistic
spin coupling coefficients, which depend on momentum, effectively
lead to the breaking of the SU(6) symmetry as discussed in
Ref.\cite{simula0}, and to a nonzero neutron charge form factor.

In this work, we focus our attention on the neutron charge mean
square radius and our aim is to explore, within a light-front
relativistic model\cite{asfbw}, the detailed dependence of
$r^2_{1n}$, obtained from $F_{1n}$, on the nucleon size that is
parametrized  by the proton charge radius ($r_p$). The proton
charge radius controls the relativistic or nonrelativistic nature
of the constituent three-quark system. For $r_p \gg m^{-1}$, the
inverse constituent quark mass, the nucleon approaches a
nonrelativistic system of quarks. On the other hand for $r_p \sim
m^{-1} \sim 1\ $fm  (the real nucleon size and constituent mass
scales), the three-quark system demands a relativistic description
of the internal quark motion. The parameters of the light-front
model are adjusted to different nucleon sizes, with several forms
of momentum component of the  wave function and  quark spin
coupling schemes. By changing the parameters we are able to shift
smoothly  from relativistic to nonrelativistic regimes. In the
nonrelativistic regime we show quite generally the cancellation
between $r^2_{1n}$ and the Foldy term contribution to the neutron
charge radius, while at the physical nucleon scale the value of
$r^2_{1n}$ is strongly dependent on the choice of quark spin
coupling scheme.

Following our previous work\cite{asfbw}, where we have studied the nucleon
electroweak form factors using different forms of relativistic
spin coupling between the constituent quarks to form the nucleon, we
use an effective Lagrangian to describe the coupling of the quark spin.
The form factor calculation keeps close contact with covariant field theory.
The starting point is the impulse approximation for the nucleon
virtual photon absorption amplitude, which is projected
on the three-dimensional null-plane hypersurface,
$x^+=x^0+x^3=0$, (see, e.g., Ref.~\cite{karmanov}). The three-dimensional
reduction is done by integrating over the individual light-front
energies ($k^-=k^0-k^3$)  in the two-loop momentum integrations of
the impulse approximation.
The relative light-front time between the particles is eliminated in favor
of the global time propagation\cite{sales00}. Then, the
momentum component  of the nucleon light-front wave function is introduced
into the remaining form of the two-loop momentum three-dimensional
integrations which define the matrix elements of the electroweak current
\cite{asfbw,tob92,pach99}. In general, the vertex function depends on the 
quark momentum variables $x_i, \vec k_{i\perp}$ and is chosen to be 
totally symmetric. We make the common assumption that the vertex function 
depend on the free three-quark light-cone energy, $M_0^2$ (Eq. 9), 
which is the simplest totally symmetric scalar function of the quark 
momentum variables.  

The effective Lagrangian for the $N-q$ coupling is written
as\cite{asfbw},
\begin{eqnarray}
{\cal{L}}_{\mathrm{N-3q}}=\alpha m_N\epsilon^{lmn}
\overline{\Psi}_{(l)}
i\tau _2\gamma _5\Psi_{(m)}^C\overline{\Psi} _{(n)}\Psi _N
+(1-\alpha)
\epsilon^{lmn}
\overline{\Psi}_{(l)}
i\tau _2\gamma_\mu \gamma _5\Psi_{(m)}^C
\overline{\Psi} _{(n)}i\partial^\mu\Psi _N
+ h.c.
\label{lag}
\end{eqnarray}
where $\tau _2$ is the isospin matrix, the color indices are
$\{l,m,n\}$ and $\epsilon^{lmn}$ is the totally antisymmetric symbol.
The conjugate quark field is $\Psi^C=C \overline{\Psi}^\top $, where
$C=i \gamma^2\gamma^0$ is the charge conjugation matrix; $\alpha$ is a
parameter to choose the spin coupling parameterization.

In Ref.~\cite{asfbw} we have tested different spin couplings for the
nucleon in a calculation of nucleon electroweak form factors. We have
found that the neutron charge form factor constrains
the relativistic  quark spin coupling schemes.
The neutron data below momentum transfers of 1 GeV/c suggested that
the scalar pair ($\alpha=1$) in the effective Lagrangian is preferred.
In that study,   Gaussian and power law momentum components of the wave
functions were used. Here we enlarge the set of momentum components of the
 wave functions to allow a wide variation of parameters, to shift from
 relativistic to nonrelativistic regimes.

This work is organized as follows. In section II, it is given a brief description
of the macroscopic and microscopic forms of the nucleon electromagnetic current
appropriate for the light-front calculations. In section III, we present the
numerical analysis of the static nucleon observables for different model
assumptions. A summary and conclusion is presented in section IV.

\section{Nucleon electromagnetic current}

\subsection{Macroscopic matrix elements}

The eletromagnetic form factors are extracted from the plus component of the
current for momentum transfers satisfying the Drell-Yan condition $q^+=q^0+q^3=0$.
The contribution of the Z-diagram is minimized in a Drell-Yan reference frame
while the  wave function contribution to the current is
maximized\cite{karmanov,tob92,pach99,brodsky,ji00}.
In particular,  the  Breit-frame is chosen, with four momentum
transfer $q=(0,\vec q_\perp,0)$,
such that $(q^+=0)$ and $\vec q_\perp=(q^1,q^2)$. The
nucleon momentum in the initial state is
$p=(\sqrt{\frac{q^2_\perp}{4}+m^2_N},-\frac{\vec q_\perp}{2},0)$ and
in the final state is given by
$p'=(\sqrt{\frac{q^2_\perp}{4}+m^2_N},\frac{\vec q_\perp}{2},0)$.

The macroscopic matrix element of the nucleon electromagnetic
current $J^+_N(q^2)$ in the Breit-frame and in  the light-front
spinor basis is given by:
\begin{eqnarray}
\langle s'|J^+_N(q^2)|s\rangle &=&\bar{u}(p',s')
\left( F_{1N}(q^2)\gamma^++ i\frac{\sigma^{+\mu}q_\mu}{2 m_N}F_{2N}(q^2)
\right) {u}(p,s)
\nonumber \\
&=& \frac{p^+}{m_N}
\langle s'| F_{1N}(q^2)+ i\frac{F_{2N}(q^2)}{2 m_N}
\vec q_\perp \cdot (\vec \sigma \times\vec n)| s \rangle \ ,
\label{jp}
\end{eqnarray}
where $F_{1N}$ and $F_{2N}$ are the Dirac and Pauli
form factors, respectively, while $\vec n$ is the unit vector along the
z-direction.

The light-front spinors are defined as:
\begin{eqnarray}
u(p,s) =\frac{\rlap\slash p+ m_N}{2\sqrt{p^+m_N}}\gamma^+
\gamma^0 \left(\begin{array}{c}
\chi^{\rm Pauli}_{s} \\0\end{array}
\right)
\ ,
\label{lf}
\end{eqnarray}
and the Dirac spinor of the instant form is given by
\begin{eqnarray}
u_{D}(p,s)=\frac{\rlap\slash p
+m_N}{\sqrt{2 m(p^0+m) }}
\left(\begin{array}{c}
\chi^{\rm Pauli}_{s} \\
0
\end{array}
\right)  \
\label{dirac}
\end{eqnarray}
which carries the subscript $D$.
The Melosh rotation is the unitary transformation between the
light-front and instant form spinors, which is given by:
\begin{eqnarray}
\left[R_{M}(p)\right]_{s's}
\ =\langle s'|\frac{p^++m-i\vec \sigma . (\vec n\times \vec p)}{
\sqrt{(p^++m)^2+p^2_\perp}}| s \rangle
\ =\overline{u}_D(p,s')u(p,s)
\ .
\label{mel}
\end{eqnarray}

\subsection{Microscopic matrix elements}

The microscopic matrix elements of the nucleon electromagnetic
current is derived from the effective Lagrangian, Eq.(\ref{lag}),
within the light-front impulse approximation which is represented by the
two-loop diagrams of figure 1\cite{asfbw}. The complete
antisymmetrization of the quark states implies that the
matrix element of the current is composed by four topologically
distinct diagrams depicted in the figure.  The matrix elements
of the electromagnetic current are calculated considering only the
process on quark 3, due to the symmetrization of the
microscopic matrix element after the factorization of the
color degree of freedom. The four distinct  current
operators $J^+_{\beta N}$, $\beta=a,b,c,d$,  are constructed from
the Feynman diagrams of figure 1a to 1d, respectively\cite{asfbw}.

The microscopic operator of the nucleon electromagnetic current,
$J^+_{N}$, is the sum of  each amplitude
 represented by the diagrams (1a) to (1d):
\begin{eqnarray}
J^+_N(q^2)=J^+_{aN}(q^2)+4J^+_{bN}(q^2)+2J^+_{cN}(q^2)+2J^+_{dN}(q^2)
\ ;
\label{mjp}
\end{eqnarray}
where the weigthing factors comes from the  identity of quarks 1 and 2,
and  another factor  2 multiplying $J^+_{bN}$ comes from
the exchange of the pairs in the initial and final  nucleons,
which gives the same matrix element as  a consequence of time reversal
and parity transformation properties.

The light-front momentum are defined as
$k^+=k^0+k^3\ , k^-=k^0-k^3 \ , k_\perp=(k^1,k^2).$
In each term of the nucleon plus component of the current, from
$J^+_{aN}$ to $J^+_{dN}$, the quark momenta are on-$k^-$-shell.
The  total plus and transverse momentum components of the intermediate
states satisfy conservation laws. Thus, the components of the momentum
$k^+_1$ and $k^+_2$ are bounded, such that $ 0< k^+_1 < p^+$ and
$0<k^+_2 <p^+-k^+_1$\cite{pach97}.

The two-loop Feynman diagram of figure 1a corresponds to
\begin{eqnarray}
\langle s'|J^+_{a N}(q^2)|s\rangle  &=& 2p^{+2}
\langle N|\hat Q_q|N\rangle
\int \frac{d^{2} k_{1\perp} dk^{+}_1d^{2} k_{2\perp} d
k^{+}_2 }{k^+_1k^+_2k^{+\ 2}_3} \theta(p^+-k^+_1)
\theta(p^+-k^+_1-k^+_2) \nonumber \\
&&{\mathrm{Tr}}\left[ (\rlap\slash k_2+m)
\left(\alpha m_N+(1-\alpha)\rlap\slash p\right)
(\rlap\slash k_1+m)\left(\alpha m_N+(1-\alpha)\rlap\slash p'\right)\right]
\nonumber \\
&&\bar u(p',s')(\rlap\slash k'_3+m))\gamma^+(\rlap\slash k_3+m)u(p,s)
\Psi (M^{'2}_0)
\Psi (M^2_0)
 \ ,
\label{j+alf}
\end{eqnarray}
where $k^2_1=m^2$ and $k^2_2=m^2$. The momentum component of the
wave function is $\Psi (M^2_0)$ and the free three-quark squared mass
is defined by:
\begin{equation}
M^2_0=p^+(\frac{k_{1\perp}^{2}+m^2}{k^+_1}+\frac{k_{2\perp}^{2}+m^2}{k^+_2}
+\frac{k_{3\perp}^{2}+m^2}{k^+_3})-{p^2_\perp} \ ,
\end{equation}
and $M^{\prime 2}_0=M^2_0(k_3\rightarrow k'_3 \ , \vec p_\perp\rightarrow
\vec p^\prime_\perp)$.
The electromagnetic
quark current operator is $\overline \Psi \hat Q_q\gamma^\mu \Psi$,
with $\hat Q_q$ the charge operator and $\Psi$ the quark field.

The other terms of the nucleon current, represented in figures 1b to 1d,
are written below:
\begin{eqnarray}
\langle s'|J^+_{b N}(q^2)|s\rangle  &=& p^{+2}
\langle N|\hat Q_q| N\rangle
\int \frac{d^{2} k_{1\perp} dk^{+}_1d^{2} k_{2\perp} d
k^{+}_2 }{
k^+_1k^+_2k^{+\ 2}_3} \theta(p^+-k^+_1)
\theta(p^+-k^+_1-k^+_2) \nonumber \\
&&\bar u(p',s')(\rlap\slash k'_3+m)\gamma^+(\rlap\slash k_3+m)
\left(\alpha m_N+(1-\alpha)\rlap\slash p\right) (\rlap\slash k_1+m)
\nonumber \\
&&\times \left(\alpha m_N+(1-\alpha)\rlap\slash p'\right)
 (\rlap\slash k_2+m)u(p,s)
\Psi (M^{'2}_0)
\Psi (M^2_0)
\ ,  \label{j+blf}
\end{eqnarray}
\begin{eqnarray}
\langle s'|J^+_{c N}(q^2)|s\rangle  &=&
p^{+2}\langle N|\tau_2\hat Q_q \tau_2| N\rangle
\int \frac{d^{2} k_{1\perp} dk^{+}_1d^{2} k_{2\perp} d
k^{+}_2 }{
k^+_1k^+_2k^{+\ 2}_3} \theta(p^+-k^+_1)
\theta(p^+-k^+_1-k^+_2) \nonumber \\
&&\bar u(p',s')(\rlap\slash k_1+m)
\left(\alpha m_N+(1-\alpha)\rlap\slash p\right)
(\rlap\slash k_3+m)\gamma^+
(\rlap\slash k'_3+m)
\nonumber \\
&&
\times \left(\alpha m_N+(1-\alpha)\rlap\slash p'\right)
 (\rlap\slash k_2+m)u(p,s)
\Psi (M^{'2}_0)
\Psi (M^2_0)
\ ,  \label{j+clf}
\end{eqnarray}
\begin{eqnarray}
\langle s'|J^+_{d N}(q^2)|s\rangle  &=& p^{+2}{\mathrm{Tr}}[\hat Q_q]
\int \frac{d^{2} k_{1\perp} dk^{+}_1d^{2} k_{2\perp} d
k^{+}_2 }{
k^+_1k^+_2k^{+\ 2}_3} \theta(p^+-k^+_1)
\theta(p^+-k^+_1-k^+_2)
\nonumber \\
&&{\mathrm{Tr}}\left[
\left(\alpha m_N+(1-\alpha)\rlap\slash p'\right)(\rlap\slash k'_3+m)
\gamma^+ (\rlap\slash k_3+m)\left(\alpha m_N+(1-\alpha)\rlap\slash p\right)
(\rlap\slash k_1+m)\right]
\nonumber \\
&&\bar u(p',s')(\rlap\slash k_2+m)u(p,s)
\Psi (M^{'2}_0)
\Psi (M^2_0)
\ .  \label{j+dlf}
\end{eqnarray}

In our study, for simplicity and as is explicit in
Eqs.(\ref{j+alf}) to (\ref{j+dlf}), the same  momentum wave
function is chosen for both $N-q$ couplings. The momentum part of
the wave function  in the microscopic matrix element of the
current is chosen from different models as we will show in the
next section.

\section{Results for Static Neutron Observables}

In this section we present our theoretical results for static
nucleon charge square radius assuming the dominance
of the lowest light-front Fock state component in the nucleon wave function
corresponding to three constituent quarks. By itself this
is a strong constraint on the static observables and
essentially the results are mostly dependent on the constituent quark mass
and scale with proton charge radius, as it will be shown.
We use a constituent quark mass value of $m=0.22 \ GeV $ from
Refs.\cite{asfbw,salme}.

Within the above assumptions, we show the effects of different
relativistic spin couplings and momentum wave functions of
constituent quarks for the neutron and proton  charge square radius.
The key point in our discussion is the detailed dependence
of
$$r^2_{1n}=6\frac{d}{dq^2}F_{1n}(q^2)$$
on the nucleon size, parameterized by the proton charge radius
($r_p$), which controls the relativistic or nonrelativistic nature
of the constituent three-quark system. The Foldy term contribution
to the neutron charge square radius is
$$ r^2_{2n}=\frac32 \frac{\mu_n}{m_N^2} \ ,$$
and the square of the neutron charge radius is the sum of both
contributions, i.e., $r^2_n=r^2_{1n}+r^2_{2n}$.

The parameters of the light-front model are changed to allow
different nucleon sizes for different forms of
momentum component of the  wave function and  quark spin coupling schemes.
By modifying the parameters a smooth shift from relativistic
to nonrelativistic regimes is obtained. The nonrelativistic regime
of the constituent quark system is characterized by $r^2_p > \ 1 \ fm^2$
(about  the inverse of the constituent quark
mass squared), and the relativistic regime by
$r^2_p < \  1\ fm ^2$  (the real nucleon size and constituent
mass scales).

The correlations between the neutron charge square
radius with magnetic moment and proton charge square radius
are investigated with a different momentum
part of the nucleon light-front wave function for each quark spin
coupling scheme.  Among the observables, the neutron charge radius plays a
special role; its correlation with the magnetic moment depends on
the quark spin coupling scheme\cite{asfbw}.

\subsection{Model wave functions}

The matrix elements of the microscopic nucleon plus component of the current,
Eqs.(\ref{j+alf}) to (\ref{j+dlf}),
are evaluated for different $N-q$ spin couplings and momentum part of the wave
function.  The different models of the momentum component of the wave function
corresponds to the choices of the  harmonic and  power-law
forms \cite{bsch,brodsky},
\begin{eqnarray}
\Psi_{\mathrm{HO}}=N_{\mathrm{HO}}\exp(-M^2_0/2\beta^2) \ \ \ , \ \ \
\Psi_{\mathrm{Power}}=N_{\mathrm{Power}}(1+M^2_0/\beta^2)^{-p}
\label{wf1}
\end{eqnarray}\\
and modified harmonic and power-law wave functions\\
\begin{eqnarray}
\Psi^{\prime}_{\mathrm{HO}}=N^{\prime}_{\mathrm{HO}}\frac{\exp(-M^2_0/2\beta^2)}
{\beta_{1}^2-M^2_0} \ \ \ , \ \ \
\Psi^{\prime}_{\mathrm{Power}}=N^{\prime}_{\mathrm{Power}}
\frac{(1+M^2_0/\beta^2)^{-p}}{\beta_{1}^2-M^2_0}
\label{wf2}
\end{eqnarray}
The normalization is determined by the proton charge.
The width parameter is $\beta$. The free mass of the three-quark
system satisfies $M_0>3m$ and $\beta_1$ has to satisfy
the constraint $\beta_{1}< 3 m$ to avoid scattering poles in the
bound state wave function of Eq.(\ref{wf2}). This property is consistent
with color confinement which prevents scattering states of three
quarks to be relevant.

The wave function models of Eq. (\ref{wf2}) are inspired in the
general form of the lowest Fock-state component of the nucleon
wave function in QCD light-front field theory where the complete
wave function  is an eigenstate of the complete mass operator
equation\cite{brodsky,pauli}. The lowest Fock state component of
the hadron wave function satisfies an effective mass operator
equation for constituent quark degrees of freedom, in which the
effective interaction contains in principle all the complexity of
QCD \cite{brodsky,pauli}. The general form of lowest Fock
component in terms of the constituent quark degrees of freedom has
the term  $(\beta_{1}^2-M^2_0)^{-1}$, where $\beta_1$ plays the
role of the mass of a bound system. Here, we use the models for
the wave function from Eq. (\ref{wf2}) just to enlarge our
possible choices of momentum components of the wave function,
while still keeping connection with the basic  QCD theory.

{}From general QCD perturbative arguments a power-law fall-off with
$p=3.5$ is predicted for $\Psi_{\mathrm{Power}}$ \cite{bsch,brodsky}.
The correlations between
static electroweak observables are not sensitive to $p$ as long as
$p>2$ \cite{asfbw,bsch} and we choose for our calculations $p=3$ in
both $\Psi_{\mathrm{Power}}$  and $\Psi^{\prime}_{\mathrm{Power}}$.
For large virtualities $\Psi_{\mathrm{Power}}\sim M_0^{-2p}$ and
$\Psi^\prime_{\mathrm{Power}}\sim M_0^{-2(p+1)}$, with $p=3$, which
is  above and below the QCD power fall-off, respectively.
As we are going to show below these different
assumptions for the wave function have only little effect in the
correlations between the static observables.

\subsection{Numerical Results}

In Figs. 2 to 5 we show results for the
correlations between static neutron charge square radius, magnetic moment and
proton radius.  Our calculations are done for
different spin couplings of quarks, i.e.  $\alpha=0$, 1/2, 1 in the
effective Lagrangian of Eq.(\ref{lag}), and momentum wave functions of
a harmonic oscillator (HO) (Gaussian) and a power-law (Power) form
($p=3$) from  Eq. (\ref{wf1}) and the modified forms from Eq. (\ref{wf2}).

The correlation of the static observables is found by variation of the
$\beta$ and $\beta_1$ parameters. In the Gaussian wave function of 
Eq.(\ref{wf1}) two limits are noteworthy, $\beta \rightarrow 0$
leads to an infinite size of the nucleon corresponding to the
nonrelativistic limit and $\beta\rightarrow \infty$ is the zero
radius limit corresponding to the strong relativistic limit. In the power-law
wave functions for $\beta\rightarrow \infty$  the relativistic limit is approached.
However in this case of $\beta\rightarrow 0$ one does not approach the nonrelativistic
limit because the typical momentum scale of the wave function is the quark mass and not
zero. The modified Gaussian and power-law wave function of Eq. (\ref{wf2}) are much more
flexible, because the nonrelativistic limit can be approached by taking  $\beta_1\rightarrow 3m$,
which corresponds to zero binding energy, and the quark system swells to infinity. Taking
$\beta$  going to infinity
in the Gaussian and in the modified Gaussian models,
the nucleon tends to zero size,
and to the extreme relativistic regime. In our results we explore
a wide range of values of  $\beta$ and $\beta_1$.

In Figure 2 results are shown for the neutron charge radius as a
function of the neutron magnetic moment for $\alpha=0$, 1/2, and 1
as well as HO, Power, modified HO and Power momentum wave
functions. The results are quite insensitive to the different
shapes of the momentum wave functions, however strongly dependent
on the quark spin coupling as we have already found in
Ref.\cite{asfbw}. The present extension to several different forms
of wave function confirms the previous findings, and moreover the
confining or nonconfining behavior of the wave function is not
important for the neutron radius as long as the magnetic moment is
fitted.  The nonconfining feature of the wave function does not
change our previous conclusion, i.e., the experimental data for
the neutron charge radius favors the scalar coupling between the
quark-pair, while the gradient  spin coupling  ($\alpha=0$)
completely disagrees with the experimental value.

As our main point is to study the neutron radius as a function of
the nucleon size, parameterized by the proton charge radius, next
we show in Figure 3 the neutron charge square radius as a function
of the proton charge square radius, for the same set of
calculations presented in Figure 2.  The different models of quark
spin couplings (for $\alpha$ equal to 0, 1/2 and 1) are shown in
Figure 3 and  represent a systematic behavior that once again is
quite independent of the form of the momentum component of the
wave function. For the chosen constituent mass ($m=0.22$ GeV) the
experimental points from \cite{brod,mur,rosen} data are within the
width for the theoretical results for the scalar coupling. The
results for the modified Gaussian and power-law wave functions are
obtained by changing either $\beta$ or $\beta_1$ in
Eq.(\ref{wf2}), which leads to a small spread of the results seen
in the figure.

The functional dependence of the individual contributions
$r^2_{1n}$ and $r^2_{2n}$ to the neutron charge square radius with
proton charge square radius  is shown in Figure 4. The
relativistic and nonrelativistic regimes are identified in the
figure. For the proton charge square radius below 3 fm$^2$, we
observe the intrinsic relativistic behavior of the quark motion
through the wide separation of the results for $r^2_{1n}$ obtained
for different quark spin coupling schemes. The nonrelativistic
regime is seen  for $r^2_p$ above 3 fm$^2$ where the calculations
tend to be flat and with magnitude which cancels to some extent
the contribution of Foldy term, $r^2_{2n}$. At the physical
nucleon size scale $r_p\sim 0.8$ fm, the  quark motion is quite
relativistic, where $r^2_{1n}$ has a sensitive dependence on the 
spin couplings, i.e. $\alpha$. The result for $r^2_{1n}$ is about 
zero for the scalar
coupling which implies in this case the dominance of the Foldy
term in the neutron charge radius. The dependence on the details
of the momentum component of the wave function is quite small.

To be complete, we present in Figure 5  the function defined by
$r^2_n(r^2_p)$ over the same range of proton sizes as in Figure 4.
The results for the neutron square charge radius show a smooth
trend in all models and spin coupling schemes from the physical
scale (relativistic regime) towards the nonrelativistic regime.
The cancellation of the Foldy term in the nonrelativistic regime
leads to the small values of $r^2_n$ for all models investigated
despite the differences in the quark spin coupling schemes and
wave functions.  We show the cancellation between $r^2_{1n}$ and
the Foldy term contribution, $r^2_{2n}$ to the neutron charge
radius in the nonrelativistic regime for the particular coupling
of Eq. (\ref{lag}), while at the physical nucleon scale the value
of $r^2_{1n}$ is strongly dependent on the choice of quark spin
coupling scheme.

\section{Summary and Conclusion}

In this work we have studied in detail the effect of the nucleon
size scale in the individual contribution of the Dirac form factor
to the neutron charge square radius, using a relativistic
light-front model with constituent quarks. The model is
constructed with different relativistic spin coupling schemes and
wave functions. The wave function parameters were adjusted to
different nucleon sizes, parameterized by the proton charge
radius, which allowed us to investigate the cancellation between
the contributions from the Foldy term ($r^2_{2n}$) and Dirac form
factor ($r^2_{1n}$)  in the neutron charge square radius that
occurs in nonrelativistic regimes\cite{isgur}.

First we extend the previous analysis of the neutron static
observables \cite{asfbw}, using more general forms of the momentum
component of the wave function, which have a nonconfining tail and
we find a model independent correlation of the Foldy term and
$r^2_{1n}$ with the proton charge radius for each spin coupling
scheme. The existence of the model independent correlation of
$r^2_{2n}$ and  $r^2_{1n}$ with the proton charge radius allowed
the study of the nucleon size scale effects on a physical ground.

Our calculations show that the cancellation between $r^2_{1n}$
and the Foldy term contribution, $r^2_{2n}$ to the neutron charge
radius happens indeed to a large extent in the nonrelativistic
regime, independent of the detailed form of quark spin coupling
scheme and wave function. The nonrelativistic regime is seen in
our calculations for the square of the proton charge radius above
3 fm$^2$ where the contributions of the Dirac form factor and
Foldy term to the neutron charge square radius
 tend to cancel. At the physical nucleon scale the value of $r^2_{1n}$ is
strongly dependent on the choice of quark spin coupling scheme,
a consequence of the intrinsic relativistic nature
of the constituent quark motion inside the nucleon.

{\bf Acknowledgments:} WRBA thanks FAPESP (Funda\c c\~ao de Amparo
\`a Pesquisa do Estado de S\~ao Paulo) for financial
support and LCCA/USP for providing computational facilities.
 TF thanks  CNPq (Conselho Nacional de Pesquisas) and FAPESP.

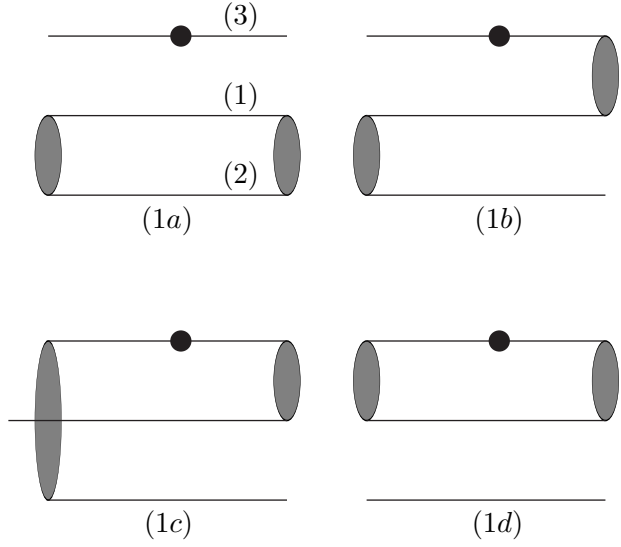
\begin{figure}[h]
\begin{center}
\vspace{5cm} \centerline{\
\begin{picture}(330,130)(-5,-130)
\GOval(30,25)(15,5)(0){.5} \GOval(120,25)(15,5)(0){.5}
\Vertex(80,70){4.0} \put(110,12){\makebox(0,0)[br]{$(2)$}}
\put(110,42){\makebox(0,0)[br]{$(1)$}}
\put(110,72){\makebox(0,0)[br]{$(3)$}} \Line(30,10)(120,10)
\Line(30,40)(120,40) \Line(30,70)(120,70)
\put(85,-5){\makebox(0,0)[br]{$(1a)$}}
\GOval(150,25)(15,5)(0){.5} \GOval(240,55)(15,5)(0){.5}
\Vertex(200,70){4.0} \Line(150,10)(240,10) \Line(150,40)(240,40)
\Line(150,70)(240,70) \put(210,-5){\makebox(0,0)[br]{$(1b)$}}
\GOval(30,-75)(30,5)(0){.5} \GOval(120,-60)(15,5)(0){.5}
\Vertex(80,-45){4.0} \Line(30,-105)(120,-105)
\Line(15,-75)(120,-75) \Line(30,-45)(120,-45)
\put(85,-120){\makebox(0,0)[br]{$(1c)$}}
\GOval(150,-60)(15,5)(0){.5} \GOval(240,-60)(15,5)(0){.5}
\Vertex(200,-45){4.0} \Line(150,-105)(240,-105)
\Line(150,-75)(240,-75) \Line(150,-45)(240,-45)
\put(210,-120){\makebox(0,0)[br]{$(1d)$}}
\end{picture}
}
\end{center}
\caption{Light-cone time-ordered diagrams for the nucleon electromagnetic 
current. The gray blob represents the spin invariant for the coupled quark
pair in the effective Lagrangian, Eq.(\ref{lag}). The black circle
in the fermion line represents the action of the current operator
on the quark. Diagram (1a) represents $J^+_{aN}$,
Eq.(\ref{j+alf}). Diagram (1b) represents  $J^+_{bN}$,
Eq.(\ref{j+blf}). Diagram (1c) represents  $J^+_{cN}$,
Eq.(\ref{j+clf}). Diagram (1d) represents $J^+_{dN}$,
Eq.(\ref{j+dlf}). } \label{fig1}
\end{figure}
\newpage
%
\begin{figure}
\setlength{\epsfxsize}{0.8\hsize} \centerline{\epsfbox{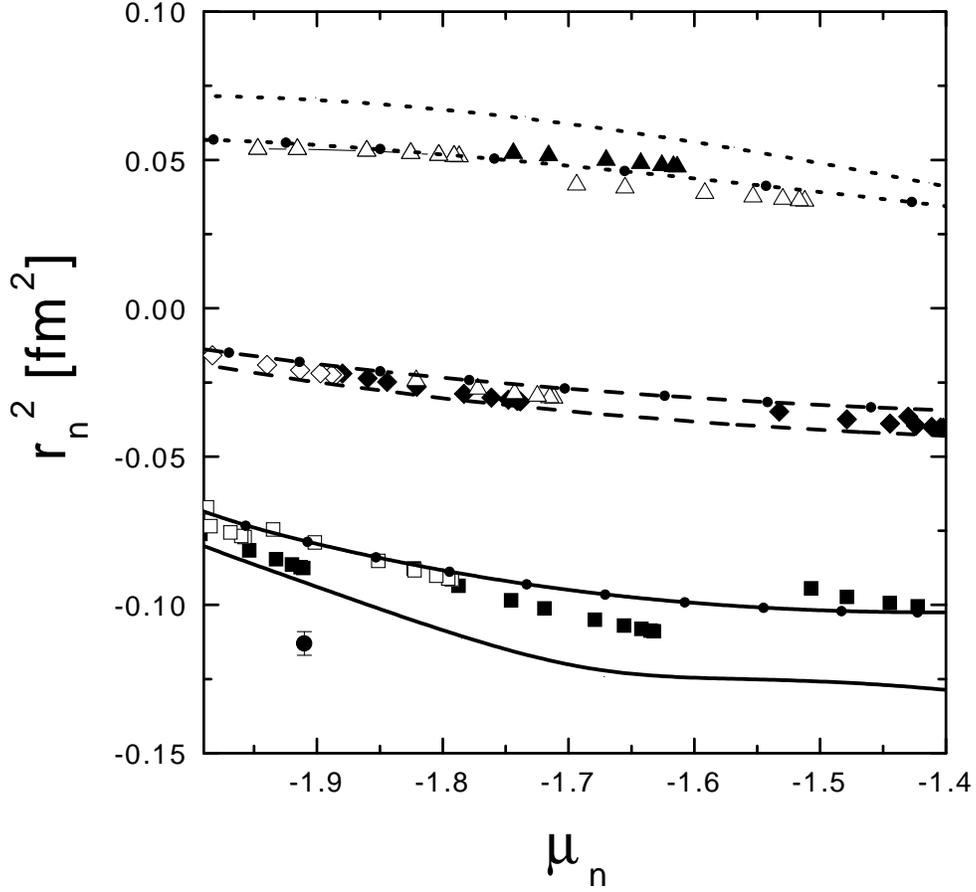}}
\caption[dummy0]{Neutron charge square radius as a function of the
neutron magnetic moment. Results for the Gaussian wave function
with $\alpha$ equal to 1 (solid line), 1/2 (dashed line) and 0
(short-dashed line). Results for the power-law wave function
with $\alpha$ equal to 1 (solid line with dots), 1/2 (dashed line with dots)
and 0  (short-dashed line with dots).
Results for modified Gaussian wave function
with $\alpha$ equal to 1 (full square), 1/2 (full diamond) and 0
(full triangle). Results for modified power-law wave function
with $\alpha$ equal to 1 (open square), 1/2 (open diamond)
and 0  (open triangle).
Experimental data from Ref.\cite{kop}. }
\label{fig2}
\end{figure}
 \begin{figure}
 \setlength{\epsfxsize}{0.8\hsize} \centerline{\epsfbox{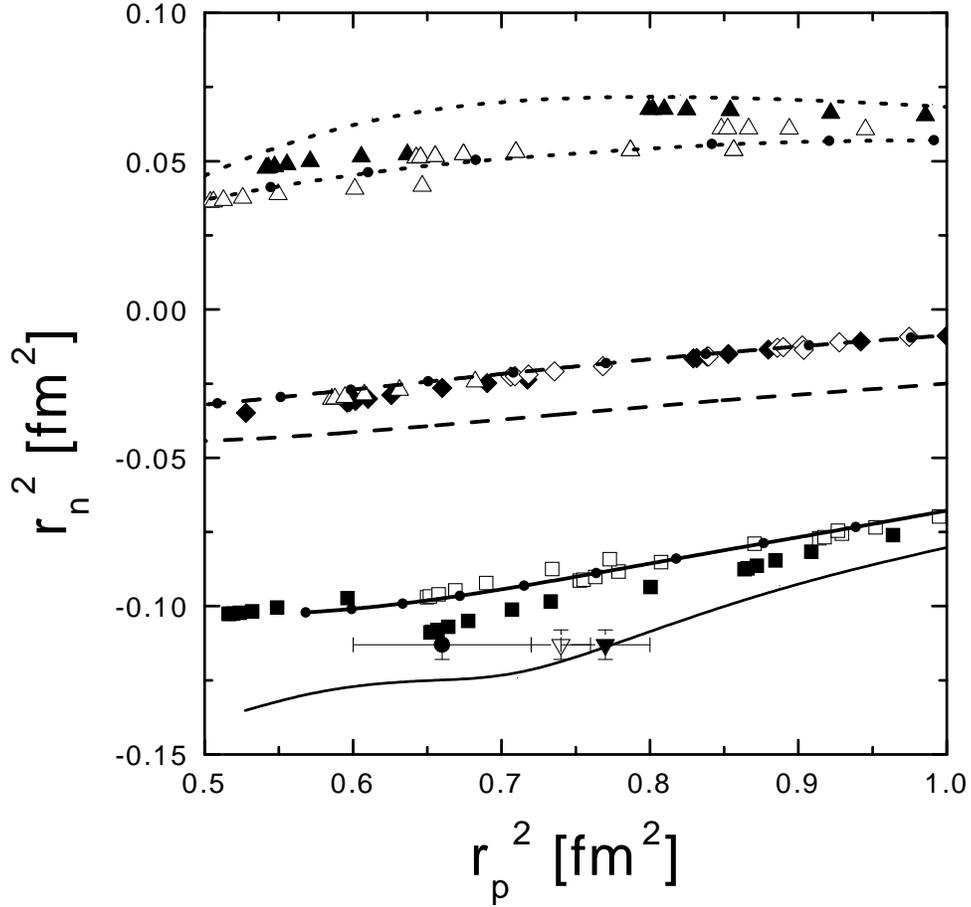}}
\caption[dummy0] {Neutron charge square radius as a function of
the proton charge square radius. Theoretical results labeled as
in figure 2. The experimental data points come from the measured
value of the neutron charge square radius\cite{kop}, and from the
experimental values of the proton charge square radius from
Refs. \cite{brod},  \cite{mur} and \cite{rosen}, which are
represented by the full circle, open inverse triangle and full
inverse triangle, respectively. } \label{fig3}
\end{figure}
\begin{figure}
 \setlength{\epsfxsize}{0.8\hsize} \centerline{\epsfbox{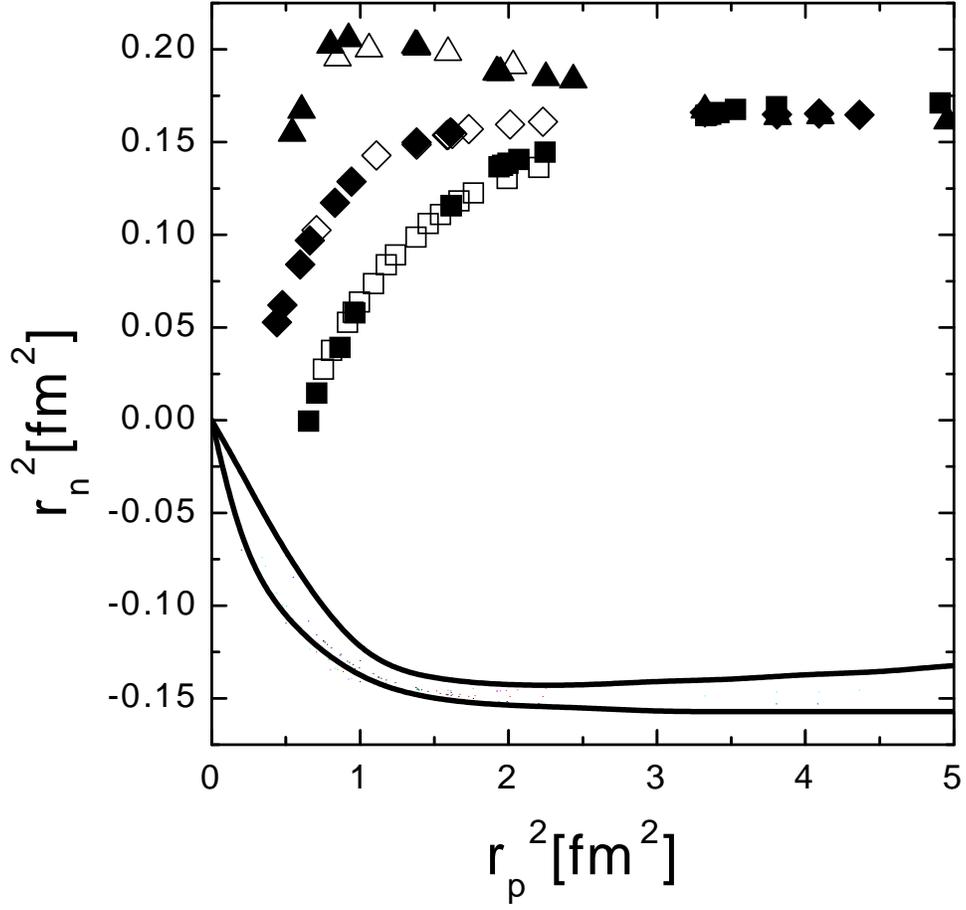}}
 \caption[dummy0]{Individual contributions to the
neutron charge radius related to the Dirac form factor $F_{1n}(q^2)$ ($r^2_{1n}$)  and from
 $F_{2n}(q^2)$  ($r^2_{2n}$ ) as a function
of the proton charge square radius. Theoretical results (squares, diamonds, 
triangles) for $r^2_{1n}$ labeled as in figure 2. Theoretical results for the 
Foldy term, $r^2_{2n}$, are bounded by the thick solid lines.  } \label{fig4}
\end{figure}
 \begin{figure}
 \setlength{\epsfxsize}{0.8\hsize} \centerline{\epsfbox{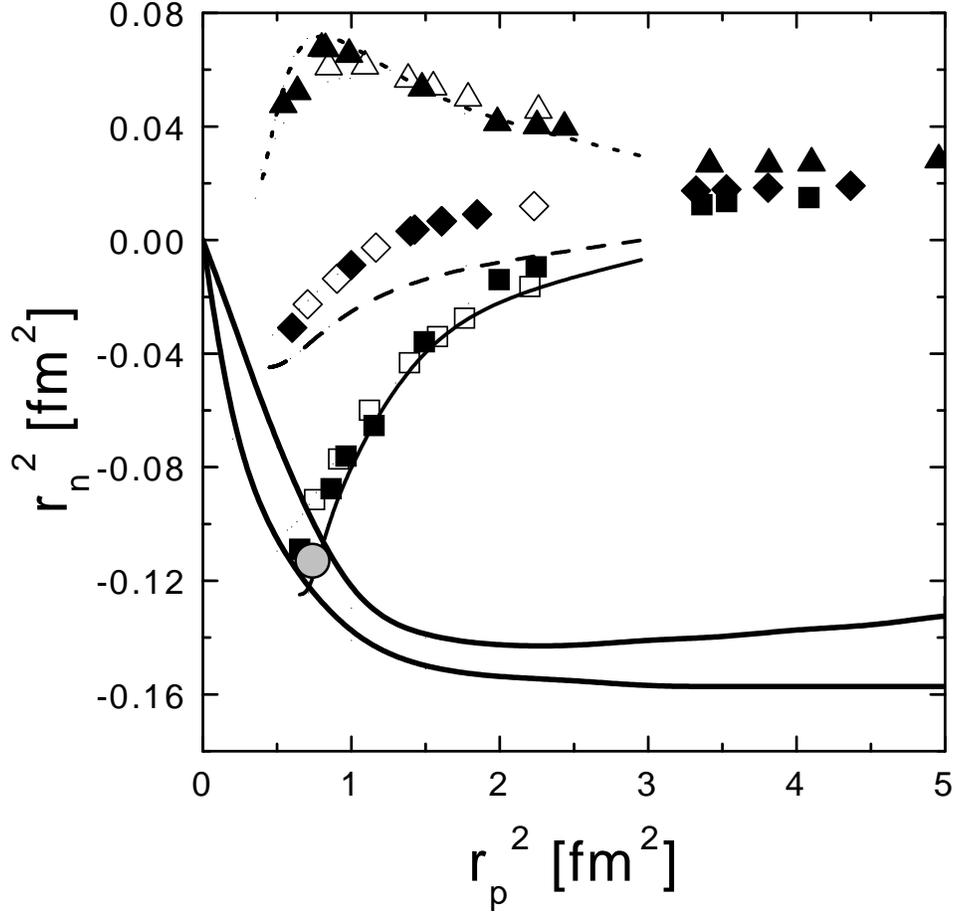}}
 \caption[dummy0]{Neutron charge square radius and
$r^2_{2n}$  as a function of the proton charge square radius.
Theoretical results for the Gaussian, modified Gaussian and
modified power-law wave functions with $\alpha$ equal to 1, 1/2
and 0, labeled as in figure 2. Experimental points included in
gray circle \cite{brod,mur,rosen}. } \label{fig5}
\end{figure}
\end{document}